\documentclass[floatfix,aps,unsortedaddress,superscriptaddress,twocolumn,sho
wpacs]{revtex4}
\usepackage[dvips]{graphicx}
\usepackage{epsfig}                                         

\begin{document}
\title{The Trojan Horse Method: an Indirect Technique in Nuclear
 Astrophysics.}

\author{A.\,M.~Mukhamedzhanov}
\affiliation{Cyclotron Institute, Texas A\&M University,
College Station, TX 77843}

\author{C.~Spitaleri}
\affiliation{DMFCI, Università di Catania, Catania, Italy and INFN -
 Laboratori Nazionali del
Sud, Catania, Italy}

\author{R.\,E.~Tribble}
\affiliation{Cyclotron Institute, Texas A\&M University,
College Station, TX 77843}

\date{\today}
                               
\begin{abstract}
The Trojan Horse (TH) method  is a powerful indirect technique that provides
 information
to determine astrophysical factors for rearrangement processes at
 astrophysically
relevant energies. A short coming for understanding the reliability of the
 technique has been 
determining the importance of nuclear and Coulomb effects on the energy
 dependence of the
yield.   
Using a simple model, we demonstrate that off-energy-shell and Coulomb effects in the entry channel and the final state nuclear interactions do not change the energy dependence of the astrophysical
factor extracted from the TH  reaction. Some examples are presented.
\end{abstract}
\pacs{26.20.+f, 24.50.+g, 25.70.Ef, 25.70.Hi}

\maketitle  

The presence of the Coulomb barrier for colliding charged nuclei
makes nuclear reaction cross sections at astrophysical energies so
small that their direct measurement in the laboratory is very
difficult, or even impossible. 
Consequently indirect techniques often are used
to determine these cross sections. The Trojan
Horse (TH) method is a powerful indirect technique which allows one to
 determine the
astrophysical factor for rearrangement reactions. 
The TH method, first suggested by Baur \cite{baur86th}, 
involves obtaining the cross section of the binary $x+ A \to b+B$ process
at astrophysical energies by measuring the two-body to three-body ($2 \to
 3$) process, $a + A \to
y + b+B$, in the
quasifree (QF) kinematics regime, where the "Trojan Horse" particle,
 $a=(x\,y)$, is accelerated
at energies above the Coulomb barrier. After penetrating through the Coulomb
barrier, nucleus $a$ undergoes breakup leaving particle $x$ to interact with
 target $A$ while
projectile $y$ flies away. From the measured $a + A \to y + b+B$  cross
 section, the energy
dependence of the binary subprocess, $x+ A \to b+B$, is determined. 

The main advantage of the
TH method is that the extracted cross section of the binary subprocess does
 not contain the
Coulomb barrier factor. Consequently the TH cross section can be used to
 determine
the energy dependence of the astrophysical factor, $S(E)$, of the binary
 process, $x+ A \to b+B$,
down to zero relative
kinetic energy of the 
particles $x$ and $A$ without distortion due to electron screening
 \cite{ass87,spit01}. 
The absolute value of $S(E)$ must be found by normalization to direct
 measurements at higher
energies. At low energies where electron screening becomes
important, comparison of the astrophysical factor determined from the TH
 method to the direct
result provides a determination of the screening potential. 

\begin{figure}[t]
\resizebox*{0.25\textwidth}{!}{\includegraphics{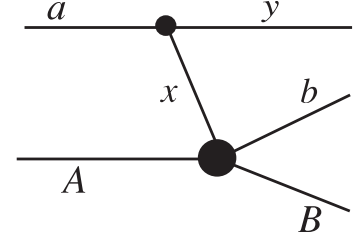}}
\caption{ The pole diagram describing the TH reaction $a+ A \to y+b+B$.}
\label{fig_pole}
\vspace{-0.6cm}
\end{figure}

Even though the TH method has been applied successfully to many direct and
 resonant processes
(see \cite{spit04} and references therein), there are still reservations
 about the reliability of the
method due to two potential modifications of the yield from off-shell
 effects and initial and final
state interactions in the TH $2 \to 3$ reaction. In the TH reaction, shown
 schematically in Fig.
\ref{fig_pole}, particle $x$ in the binary subprocess $x+ A \to
b+B$ is virtual (off-energy-shell). 
In the standard analysis, the
virtual nature of $x$ is neglected and the plane wave approximation is used \cite{spit01,typ03}. 
Here we address, for the first time, the reliability of this assumption. We
 also consider
scattering between particles
$a$ and $A$ in the initial channel of the TH reaction and the dominance of
 the
QF mechanism. Note that the TH reaction is a many-body process (at least
 four-body) and its
strict analysis requires many-body techniques. However some important
 features of the TH
method can be addressed in a simple model.  

First the notation for the subsequent discussion is introduced. $E_{ij}(=
k_{ij}^{2}/2\,\mu_{ij}$)
and ${\bf{\rm k}}_{ij}$ are the relative energy and momentum of the
real particles $i$ and
$j$, $\mu_{ij}=m_{i}\,m_{j}/(m_{i} + m_{j})$, 
where $m_{i}$ is the mass of particle $i$, and ${\rm {\bf p}}_{xj}$ is the
relative momentum of the virtual particle $x$ and particle $j$. 
The internal particle $x$ of the diagram shown in Fig. \ref{fig_pole} is
virtual, i.e. $E_{x}
\not= p^{2}_{x}/2\,m_{x}$. From energy-momentum conservation laws in the
three-ray vertex $a \to x + y$ and the four-ray vertex $x + A \to b + B$, we
get $\sigma_{x}=p_{xy}^{2}/(2\,\mu_{xy}) +
\varepsilon_{xy}^{a}=p_{xA}^{2}/(2\,\mu_{xA}) - E_{xA}$,
where $\varepsilon_{xy}^{a}= m_{x} + m_{y} - m_{a}$ is the binding energy
for the virtual decay $a \to x+ y$. Thus in a TH reaction 
$p_{xA}^{2}/(2\,\mu_{xA}) >E_{xA}=k_{xA}^{2}/(2\,\mu_{xA})$ holds. 
The reaction amplitude corresponding to the diagram
is given by (all particles are assumed, for simplicity, to be spinless)
$M_{p}= M_{2 \to 2}^{(HOF)}\,\varphi_{xy}(p_{xy})$.  
Here $M_{2 \to 2}^{(HOF)}(\sigma_{x},E_{xA},z)$ is the half-off-energy-shell
(HOF) reaction amplitude for the direct binary subprocess $x+ A \to b + B$
which depends on the additional variable $\sigma_{x}$
due to the virtual nature of particle $x$. Also 
$z= {\hat {\rm {\bf{p}}}}_{xA} \cdot {\hat {\rm {\bf{k}}}}_{bB},\;$ with 
${\hat {\rm {\bf{k}}}}={\rm {\bf{k}}}/k$.  
The Fourier transform of the $s$-wave bound-state wave function for
$a=(x\,y)$
can be written as $\varphi_{xy}( p_{xy})=-W_{xy}( p_{xy})/(\sigma_{x})^{1-
\eta_{xy}}$, where $\eta_{xy}$ is the Coulomb parameter of the bound 
state $(xy)$ and $W_{xy}$ is the amplitude for the virtual decay $a \to x +
 y$, which is regular
at the singular point $\sigma_{x}=0$. This singularity is a branch point
 for the charged 
particles $x$ and $y$. For $a=d=(p\,n)$, $\varphi_{xy}( p_{xy})$ has a
 pole at
$\sigma_{x}=0$, corresponding to the real (on-energy-shell (ON)) particle
 $x$. Hence the name 
pole diagram for Fig. \ref{fig_pole} \cite{chewlow}. The modulus of the amplitude
$|M_{p}|$ of the pole diagram has a maximum at $p_{xy}=0$ called the QF
 peak and the
condition corresponding to $p_{xy}=0$ is called QF kinematics. The QF peak
 is a trace
of the pole located at $p_{xy}^{2} <0$. 
The QF kinematics are ideal for the TH method since small $p_{xy}$
corresponds to large separation distance between particles $x$ and $y$,
 thereby allowing
particle $y$ to be treated as a spectator. 

By dropping all the terms containing particle $y$ and its interaction with
 other nuclei, the half-off-shell post-form amplitude of the direct binary reaction, $A(x, b)B$, can be extracted from the exact amplitude for the TH reaction, $A(a, y\,b)B$. The result is given by
\begin{equation}
M_{2 \to 2}^{(HOF)}(\sigma_{x},E_{xA},z)  \sim < \psi
 _{bB}^{(-)}\varphi_b \varphi
_B |{\hat O}|\varphi_{x}\varphi_{A}{\bf {\rm p}}_{xA}>. 
\label{binampl1} 
\end{equation}
Here, $\psi _{bB}^{(-)}$ is the distorted wave in the final state of the
 binary process,
$\varphi_i$ is the bound state wave function of nucleus $i$, 
${\hat O}= \Delta V_{bB}\,[1 +  G_{xA}^{(+)}\,\Delta V_{xA}]$ is the
 transition operator,
$\Delta V_{ij}= V_{ij} -U_{ij}$, $\,V_{ij}$ ($U_{ij}$) is the interaction
 (optical) potential
between nuclei $i$ and $j$ and $G_{xA}^{(+)}$ is the Green's function of the
system $x+ A$ (or $b+ B$). The half-off-shell amplitude contains the off-shell plane wave $|{\bf {\rm p}}_{xA}>$ which describes the relative motion of the virtual particle $x$ and $A$ in the initial
channel of the binary reaction rather than the distorted wave describing the
initial state in the on-shell amplitude. Hence the half-off-shell amplitude does not contain a Coulomb barrier factor. 
First we estimate the contribution from the transition operator $[\Delta
V_{bB}\,G_{xA}^{(+)}\,\Delta V_{xA}]$ to $M_{2 \to 2}^{(HOF)}$.  
Consider two reactions previously analyzed using
the TH method, ${}^{6}{\rm Li}(d,\alpha){}^{4}{\rm He}$ ($Q_{2 \to 2}=
 21.64 $ MeV) \cite{spit01} and ${}^{7}{\rm Li}(p,\alpha){}^{4}{\rm He}\,$ ($Q_{2 \to 2} = 16.56$ MeV) \cite{lat01}. The first reaction was treated as a deuteron transfer and the second as a triton transfer. Nuclei $A$ and $b$ are bound states $A=(B\,t)$ and $b=(x\,t)$ with constituent particles $x,\,t,\,B$, where $t$ is the transferred
particle. In this simple model the half-off-shell amplitude takes the form
(all particles are assumed to be spinless)
\begin{equation}
M_{2 \to 2}^{(HOF)}(\sigma_{x},E_{xA};z) \sim < \psi
 _{bB}^{(-)}\varphi_{xt}|{\hat
O}|\varphi_{Bt}{\bf {\rm p}}_{xA}>. 
\label{binampls1} 
\end{equation}
The bound state wave functions have been approximated by their tails,
$\Delta V_{bB} \approx V_{Bt}$ and the zero-range approximation (ZRA) has been
used for $V_{Bt}\,\varphi_{Bt}$. The ZRA is good enough to 
determine the energy dependence of the cross section. At $E_{xA} < 1$ MeV, 
only the Coulomb interaction needs to be included in the transition operator $\Delta V_{xA}$ and in the Green's function. 
Also at low $E_{xA}$ the dominant contribution comes only from the $s$-wave
in the channel $x+ A$ due to the large $Q_{2 \to 2}$ value for both reactions. Consequently the
$z$ dependence of $M_{2 \to 2}^{(HOF)}$ can be neglected. 
The contribution from the transition operator $[\Delta V_{bB}G_{xA}^{(+)}\Delta V_{xA}]$ is negligible for both reactions as is the case for the on-shell processes. It constitutes a background
for the DWBA with the transition operator given by $\Delta V_{bB}$. 

\begin{figure}[t]
\resizebox*{0.45\textwidth}{!}{\includegraphics{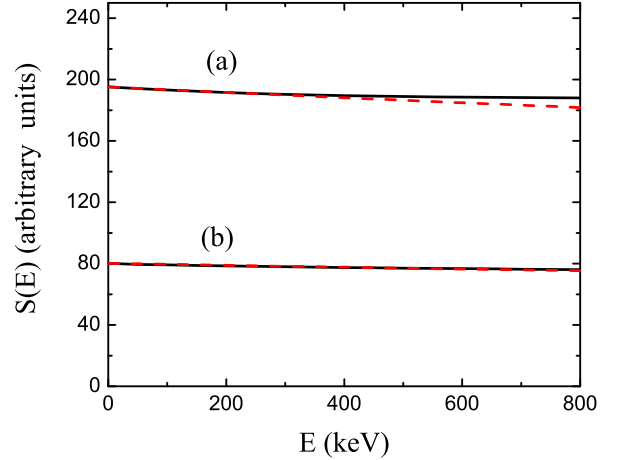}}
\vspace{-0.3cm}
\caption{(Color online). Energy dependence ($E \equiv E_{xA}$) of the half-off-shell (red dashed line) 
and on-shell (black solid line)
astrophysical factors for (a) the ${}^{7}{\rm Li}(p,\alpha){}^{4}{\rm He}$
reaction ($a=d,\,x=p,\,y=n$); (b) the 
${}^{6}{\rm Li}(d,\alpha){}^{4}{\rm He}\,$ reaction ($a={}^{6}{\rm
 Li},\,x=d,\,y=\alpha$).}
\label{fig_halfonSfctr}
\vspace{-0.1cm}
\end{figure}

The on-shell amplitude to be compared with the half-off-shell one is 
\begin{equation}
M_{2 \to 2}^{(ON)}(E_{xA})  \sim < \psi
 _{bB}^{(-)}\varphi_{xt}|\Delta
V_{bB}|\varphi_{Bt}\psi_{xA}^{(+)}>. 
\label{binampl2} 
\end{equation}  
Here, $\psi_{xA}^{(+)}$, the distorted wave describing the scattering of
real particles $x$ and
$A$ in the initial channel, can be
approximated by the pure Coulomb scattering wave function at low energies.
Consequently, 
$\psi_{xA}^{(+)}= 
N_{xA}\,exp(i\,{\rm {\bf k}}_{xA} \cdot {\rm {\bf r}}_{xA})\,{}_{1}F_{1}$,
 where
$N_{xA}$ is the Gamow normalization factor and
${}_{1}F_{1}$ is the hypergeometric function,
which has a very weak energy dependence at small $k_{xA}$. The only
difference in the energy dependence between the half-off-shell and on-shell astrophysical factors comes from the 
the use of the half-off-shell plane wave in the initial state in Eq.
(\ref{binampls1}) 
and the on-shell distorted wave in Eq. (\ref{binampl2}).
From Fig. \ref{fig_halfonSfctr} it is clear that the results, which were calculated for QF kinematics, clearly justify
the TH method.  The energy dependence of the half-off-shell and on-shell
astrophysical factors at low energies are practically identical. Since only the energy dependence is of interest, 
the half-off-shell result in Fig. \ref{fig_halfonSfctr} 
has been normalized to the on-shell one at an energy 
$E_{xA}= 1$ keV for ease of comparison. 

\begin{figure}[t]
\resizebox*{0.45\textwidth}{!}{\includegraphics{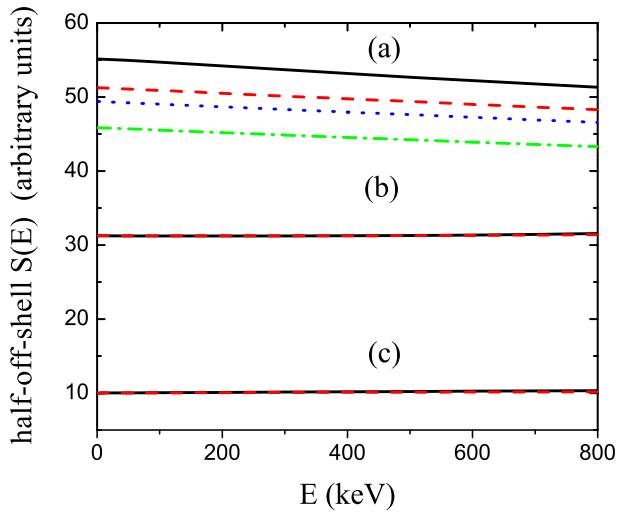}}
\vspace{-0.2cm}
\caption{(Color online). (a) Energy dependence ($E \equiv E_{{d}^{6}{\rm Li}}$) of the half-off-shell astrophysical factor for the
${}^{6}{\rm Li}(d, \alpha){}^{4}{\rm He}$ reaction as a function of
 $\sigma_{d}$:
$\sigma_{d}=0$- black solid line (on-shell kinematics);
$\sigma_{d}=\varepsilon_{d \alpha}^{{}^{6}{\rm Li}}$- red dashed line (QF
 kinematics);
$\sigma_{d}=1.5\,\varepsilon_{d \alpha}^{{}^{6}{\rm Li}}$ - blue dotted
 line; 
$\sigma_{d}=2.5\,\varepsilon_{d \alpha}^{{}^{6}{\rm Li}}$ - green
 dashed-dotted line.
(b) Comparison of the energy dependence of the half-off-shell astrophysical
factor for the
${}^{6}{\rm Li}(d,\alpha){}^{4}{\rm He}$ reaction determined from the 
${}^{6}{\rm Li}({}^{6}{\rm Li},\alpha\,\alpha){}^{4}{\rm He}$ TH reaction
with (black solid line) and
without (red dashed line) inclusion of the initial ${}^{6}{\rm Li}-{}^{6}{\rm Li}$ Coulomb interaction. (c) The same as in (a) but with the final-state interaction described by the Ali-Bodmer potential: 
$\sigma_{d}=0$- black solid line (on-shell kinematics);
$\sigma_{d}=\varepsilon_{d \alpha}^{{}^{6}{\rm Li}}$- red dashed line (QF
 kinematics). In both cases (b) and (c) the red dashed line is normalized to the black solid line at $E= 1$ keV.}
\label{fig_virtualitydependence}
\vspace{-0.3cm}
\end{figure}
Another way to determine the effect of the virtual nature of the transferred
particle is to vary the
parameter $\sigma_{x}$.  In Fig. \ref{fig_virtualitydependence} (a) the
energy dependence of the half-off-shell astrophysical factor for the 
${}^{6}{\rm Li}(d,\alpha){}^{4}{\rm He}$ reaction ($x=d; A={}^{6}{\rm Li};
b=B=\alpha$) is plotted for different values of 
$\sigma_{d}$. Note that $\sigma_{d}=0$ corresponds to the on-shell
deuteron. The deuteron becomes farther from the energy-shell 
as $\sigma_{d}$ grows. 
It is clear from Fig.
\ref{fig_virtualitydependence} that the size of the $S$ factor changes but
the energy dependence does not change. This justifies the 
procedure of disregarding the virtual nature of the entry particle in the
analysis of the binary
process $A(x,b)B$ \cite{spit01,typ03}.  
If the amplitude of the TH process $A(a,y\,b)B$ is 
approximated by the pole amplitude $M_{p}$, as is done in the TH method,
then the amplitude of the binary subprocess can be trivially singled out. 
This approximation ignores the interaction between particles 
$a$ and $A$ in the entry channel. 
Inclusion of the Coulomb interaction in the $a-A$ channel is a way to gauge
the validity of this approximation.  Note that including the
Coulomb-nuclear interaction in the exit channel does not change the final
result. The TH reaction amplitude with Coulomb scattering in the entrance channel is given by
\begin{equation}
M^{(TH)} \sim < \psi_{yF}^{(-)}\psi _{bB}^{(-)}\varphi_{xt}|\Delta
V_{bB}|\varphi_{Bt}\,\varphi_{xy}\,\psi_{aA}^{(C)(+)}>, 
\label{THampl1} 
\end{equation}
where $\psi_{yF}^{(-)}$ is the distorted wave describing the scattering of
$y$ in the center-of-mass of the system $F=b + B$. As an example, consider
the TH reaction ${}^{7}{\rm Li}(d,n\,\alpha){}^{4}{\rm He}$, where 
$a=d=(pn),\,A={}^{7}{\rm Li}$. The bound state wave function of $d$ is approximated by its tail. The Coulomb interaction between 
$d-{}^{7}{\rm Li}$ modifies the behavior of the amplitude of the
TH reaction. For simplicity, the final state scattering wave functions are
replaced by plane waves.
To single out the amplitude of the binary process ${}^{7}{\rm
Li}(p,\alpha){}^{4}{\rm He}$ in
QF kinematics, the 
$M^{(TH)}$ is divided by the factor
\begin{equation}
R=<\varphi_{xy}({\rm {\bf k}}_{yF}-(m_{y}/m_{a})\,{\rm {\bf
p}}_{aA})|\psi^{(C)(+)}_{{\rm
{\bf k}}_{aA}}({\rm {\bf p}}_{aA})>, 
\label{Rfact} 
\end{equation}
where integration is performed
over ${\rm {\bf p}}_{aA}$ and $ \psi^{(C)(+)}_{{\rm {\bf k}}_{aA}}
({\rm {\bf p}}_{aA})$ is the Fourier component of the $a-A$ Coulomb scattering wave function. This factor determines the behavior of  
$M^{(TH)}$ near the singularity $\sigma_{x}=0$. Note that for the results
presented in Fig. \ref{fig_halfonSfctr} the half-off-shell astrophysical
factor has been calculated at a fixed $p_{xy}=0$ corresponding to QF 
kinematics of the TH $2 \to 3$ reaction. $E_{xA}$ becomes
a unique function of $E_{aA}$. In practice TH experiments are carried out at a fixed energy, $E_{aA}$, while
$p_{xy}$ is allowed to vary around $p_{xy}=0$. For example, to cover the
energy region $0 < E_{p{}^{7}{\rm Li}} \leq 800 $ keV for the ${}^{7}{\rm Li}(d,\,n\alpha){}^{4}{\rm He}$
reaction at energy $E_{d{}^{7}{\rm Li}} = 4.20$ MeV, it is sufficient to
vary the factor
$\sigma_{p}$ by  $7\%$ around the QF value $\sigma_{p}= 2.224$ MeV. 

To test the importance of the Coulomb interaction in the initial state, the
energy dependence of the half-off-shell astrophysical factor for the binary ${}^{6}{\rm Li}(d,\alpha){}^{4}{\rm He}$ reaction, determined from the TH reaction ${}^{6}{\rm Li}({}^{6}{\rm Li},\alpha\,\alpha){}^{4}{\rm He}$ at $E_{{}^{6}{\rm Li}{}^{6}{\rm
 Li}}=3.14$ MeV, the TH reaction amplitude (Eq. \ref{THampl1}) was calculated with and without the initial Coulomb interaction term. The results, as seen in Fig. \ref{fig_virtualitydependence} (b), show that the energy dependence of both astrophysical factors is nearly identical. Thus including
the Coulomb interaction in the initial state does not affect the energy
dependence of the astrophysical factor determined from the TH reaction.
Similar calculations done for 
$S(E_{p{}^{7}{\rm Li}})$ for the binary ${}^{7}{\rm
 Li}(p,\alpha){}^{4}{\rm He}$ reaction, determined from the TH reaction ${}^{7}{\rm Li}(d,\,n\,\alpha){}^{4}{\rm He}$ at $E_{d{}^{7}{\rm
 Li}}=4.03$ MeV, show the same result. As also seen from Fig. \ref{fig_virtualitydependence} (c), the final state interaction (here described by the Ali-Bodmer $\alpha-\alpha$ potential) does not affect the energy 
behavior of the on-shell and half-off-shell astrophysical factors for 
the binary process ${}^{6}{\rm Li}(d,\alpha){}^{4}{\rm He}$.

Understanding TH reactions proceeding through a resonant binary subprocess 
is simpler than for the direct case. The TH pole mechanism for the direct process near the QF peak is dominated by the singularity (pole for $a=d$ or branch point for charged particles $x$ and $y$) in the bound state wave function $\varphi_{xy}(p_{xy})$. In a resonant reaction, the mechanism depends on
two singularities, one that occurs in the bound state wave function in the
 $p_{xy}$ plane and the
other that is due to the resonant pole of the amplitude of the binary
 subprocess $A(x,b)B$ in the
$E_{xA}$ plane. The
TH $2 \to 3$ reaction proceeding through the resonance in the subsystem
 $x+A$ is very similar
to stripping $(d,p),\,(d,n)$ or $\alpha$-transfer $({}^{6}{\rm Li},d)$
 reactions populating
resonant states \cite{bun71}. 
\begin{figure}[t]
\resizebox*{0.45\textwidth}{!}{\includegraphics{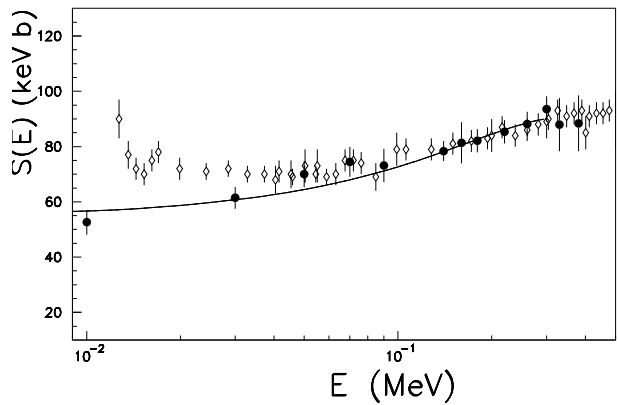}}
\caption{$S(E)$ for the ${}^{7}{\rm Li}(p,\alpha){}^{4}{\rm He}$ reaction.  The solid curve is the polynomial fit to the data (full circles) obtained from the TH  ${}^{7}{\rm
Li}(d,\alpha\,n){}^{4}{\rm He}$ \cite{lat01} reaction.
Open cirles are the data from the direct measurement which include electron screening effects \cite{engstler92}.}
\label{fig_p7Li}
\end{figure}  
The difference is that in the TH process we are interested in the decay of
the intermediate
resonances into the channel $b+ B$ which differs from the entry channel $x +
 A$. The
half-off-shell $s$-wave resonant amplitude is given by  
\begin{eqnarray}
 M^{(HOF)(R)} ( k_{bB} ,\, p_{xA} ; E) = - 
2\,\pi\,\sqrt {\frac{1}{{\mu _{bB} k_{bB} }}}\,e^{i\delta _{f0 }(k_{bB} 
)} \nonumber\\
\times \frac{{\sqrt {\Gamma _{bB} (E_{bB} )} \,w_{0 } (p_{xA} 
,k_{xA(R)} )}}{{E_{xA}  - E_{xA}^{(R)} }}. 
\label{resreact11}
\end{eqnarray}
Here, $\Gamma_{bB} (E_{bB})$ is the partial width of the resonance 
$F^{*}$ in the channel $b + B$,  $k_{xA(R)} =
 \sqrt{2\,\mu_{xA}\,E_{xA}^{(R)}}$, 
$E_{xA}^{(R)}$ is the resonance energy and $\,\delta _{f0}$ is the
 nonresonant (potential)
scattering phase shift of particles $b$ and $B$ in the final
state. The off-shell $s$-wave form factor for the vertex $x + A \to F^{*}$
is given by 
$w_{0 } (p_{xA},k_{xA(R)})=[E_{xA}^{(R)}  -
 p_{xA}^{2}/(2\,\mu_{xA})]\,\psi _{n0
}^{(R)}(p_{xA})$,  where 
$\psi _{n0 }^{(R)}(p_{xA})=<j_{0 }(p_{xA}\,r)|\psi _{n0 }^{(R)}(r)>$
is the Fourier
transform of the Gamow radial wave
function $\psi _{n0 }^{(R)}(r)$, $\;j_{0 } (p_{xA}\,r)$ is the $s$-wave 
spherical Bessel function and $n$ is the principal quantum number. 
Due to factorization of the off-shell form factor in the half-off-shell
resonant amplitude, the ratio of the half-off-shell to on-shell 
astrophysical factors is $S^{(HOF)(R)}/S^{(ON)(R)} = k_{xA}\,w_{0}(p_{xA} 
,k_{xA(R)} )/[(e^{-2\pi\,\eta_{xA}})\Gamma_{xA}(E_{xA})]$, where $\eta_{xA}$
 is the
Coulomb parameter. Taking into account that 
$\Gamma_{xA}(E_{xA}) \sim k_{xA}\,P_{0}(E_{xA})$, where $P_{0}(E_{xA})$ is
the
barrier penetrability, we conclude that the energy dependence of the
 half-off-shell
and on-shell astrophysical factors at low energies is the same. Note that if
the Coulomb interaction in the initial state is included, the
half-off-shell resonant reaction amplitude should be divided by the same
factor
$R$ as for direct processes. 
\begin{figure}[t]
\resizebox*{0.45\textwidth}{!}{\includegraphics{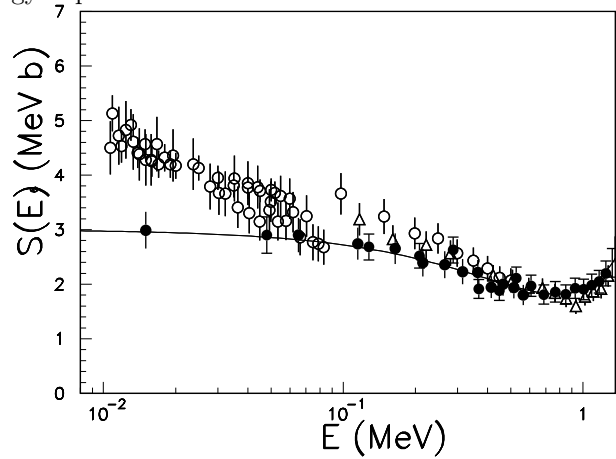}}
\caption{$S(E)$ for  ${}^{6}{\rm Li}(p,\alpha){}^{3}{\rm He}$ $\,S(E_{p
 {}^{6}{\rm Li}})$ 
from the TH reaction ${}^{6}{\rm Li}(d,\alpha\,n){}^{3}{\rm He}$ (solid
 circles) \cite{tum03}
compared to direct data (open triangles \cite{elwyn79} and open
circles \cite{engstler92}). The line shows the result of a second order
 polynomial fit to
the TH reaction data \cite{tum03}.}
\label{fig_halfonSfctr1}
\end{figure}
          
A simple model has been used to compare the energy behavior of the direct
 and resonant
half-off-shell and on-shell astrophysical factors. The intention was not to
 reproduce experimental
data, which requires a more sophisticated approach, but to demonstrate that
 the energy
dependence of the half-off-shell and on-shell astrophysical factors are
 nearly identical when
analyzed in the same model.  Validating this makes it clear why the TH
 method is such a
powerful indirect technique for nuclear astrophysics. 
The power of the TH method is seen in Figs \ref{fig_p7Li} and \ref{fig_halfonSfctr1}. In both cases the energy dependence of the astrophysical factor determined from the TH reaction nicely reproduces the energy
dependence of the astrophysical factor obtained from direct measurements
at higher energies. At lower energies the TH method provides the astrophysical factor between bare nuclei while the direct data are distorted by electron screening.

This work was supported in part by the U.\,S. DOE under Grant No.\@
 DE-FG02-93ER40773 and the
U.\,S. NSF under Grant No. \@ PHY-0140343.

\end{document}